\documentclass{article}

\usepackage{arxiv}

\usepackage[utf8]{inputenc} 
\usepackage[T1]{fontenc}    
\usepackage{hyperref}       
\usepackage{url}            
\usepackage{booktabs}       
\usepackage{amsfonts}       
\usepackage{nicefrac}       
\usepackage{microtype}      
\usepackage{lipsum}
\usepackage{cite}
\usepackage{amsmath,amssymb,amsfonts}
\usepackage{algorithmic}
\usepackage{graphicx}
\usepackage{textcomp}
\usepackage{epstopdf}
\usepackage{xcolor}
\usepackage{subfig}
\usepackage{siunitx}
\usepackage{textcomp}
\usepackage{amsmath}

\title{Graphene-Based Multifunctional Three-Port Far-Infrared Components}

\author{
  Victor Dmitriev\thanks{Use footnote for providing further
    information about author (webpage, alternative
    address)---\emph{not} for acknowledging funding agencies.} \\
 Department of Electrical Engineering\\
  Federal University of Para\\
  66075-900, Belem, Para, Brazil. \\
  \texttt{victor@ufpa.br} \\
   \And
Geraldo Melo\\
  Federal Rural University of Amazonia\\
68700-030, Campus Capanema, Para, Brazil.\\
    \texttt{geraldo.melo@ufra.edu.br} \\
  \AND
  Wagner Castro\\
  Institute Cyberspace\\
  Federal Rural University of Amazonia\\
 66.077-830, Belem, Para, Brazil.\\
    \texttt{wagner.ormanes@ufra.edu.br} \\
}

\begin{document}
\maketitle

\begin{abstract}
Two graphene-based T-shaped multifunctional components for THz and Far-Infrared regions are proposed and analyzed. The first component can serve as a divider, a switch and a dynamically controllable filter. This T-junction presents a circular graphene resonator and three graphene waveguides with surface plasmon-polariton waves connected frontally to the resonator. The resonator can be adjusted to work with dipole, quadrupole or hexapole modes. The graphene elements are deposited on a $SiO_{2}$ (silica) and $Si$ (silicon) two-layer substrate. The dynamical control and switching of the component is provided by the electrostatic field which defines the graphene Fermi energy. Numerical simulations show that the first component in the division regime (which is also the regime ON) has the transmission coefficient -4.3 dB at the central frequency for every of the two output ports and the FWHM is 9.5$\%$. In the OFF regime, the isolation of the two output ports from the input one is about -30 dB. The second component is a T-junction without  resonator which fulfils  the function of the divider-switch in more than octave frequency band.

\end{abstract}


\section{Introduction}
Switch is a key component of modern digital technology. Frequently, it is also used as a modulator.
The power divider (splitter) is another important component in the wave-guiding systems. In THz and infrared regions, several types of the switches and dividers have been described in literature \cite{Giani, L, Design, Numerical, Beam, splitter}. Filters also play an important role in communication and other systems. 

Two-dimensional  materials, especially graphene, have a great potential for their use in photonics due to high interaction with electromagnetic radiation in THz and infrared bands. Many   components for telecommunication systems based on graphene have been suggested recently, such as filters, dividers, sensors, switches, modulators, among others \cite{Geim, Slot, Bandpass, filtro, chave, sensor, modulator, divisor}.

Excitation of plasmon resonances in graphene, generating surface plasmon polariton (SPP) waves, provides a better performance  compared to the noble metals, making graphene an excellent material for many waveguiding components \cite{Graphene, ultrashort, wagner}. Graphene  has another advantage, namely, a possibility of  control by external electric field, and moreover, it can be fulfilled dynamically. In addition, the high confinement of plasmonic waves in graphene favours creation of more compact and faster components with high transmission rates, good bandwidth and relatively low losses, allowing realization of devices that can operate over a wide range of frequencies \cite{ultra, Four}.

Further reduction of losses and  dimensions of the systems can be achieved by combination of several functions in one device. In this work, we suggest a multifunctional component based on a graphene resonator and three waveguides. The circular resonator provides  dipole, quadrupole or hexapole resonant modes \cite{Nikitin, Zhang, filter,90, Mid}. A combination of these modes allows one  to provide simultaneously the functions of a controllable filter, switch and divider. In a common realization of such optical circuits, one would use a serial connection of three separated elements, namely, a filter, a switch, and a divider. In our case, a unique element fulfils all these functions, and what is more, this element provides a very effective dynamical control of its central frequency. Another suggested element is two-functional, namely, it is a wideband non-resonant divider-switch. These components can be used in THz and far-infrared regions. 
\section{Problem Description}
\label{sec:examples}
\begin{figure}[htbp]
\centering
\includegraphics[width=\linewidth]{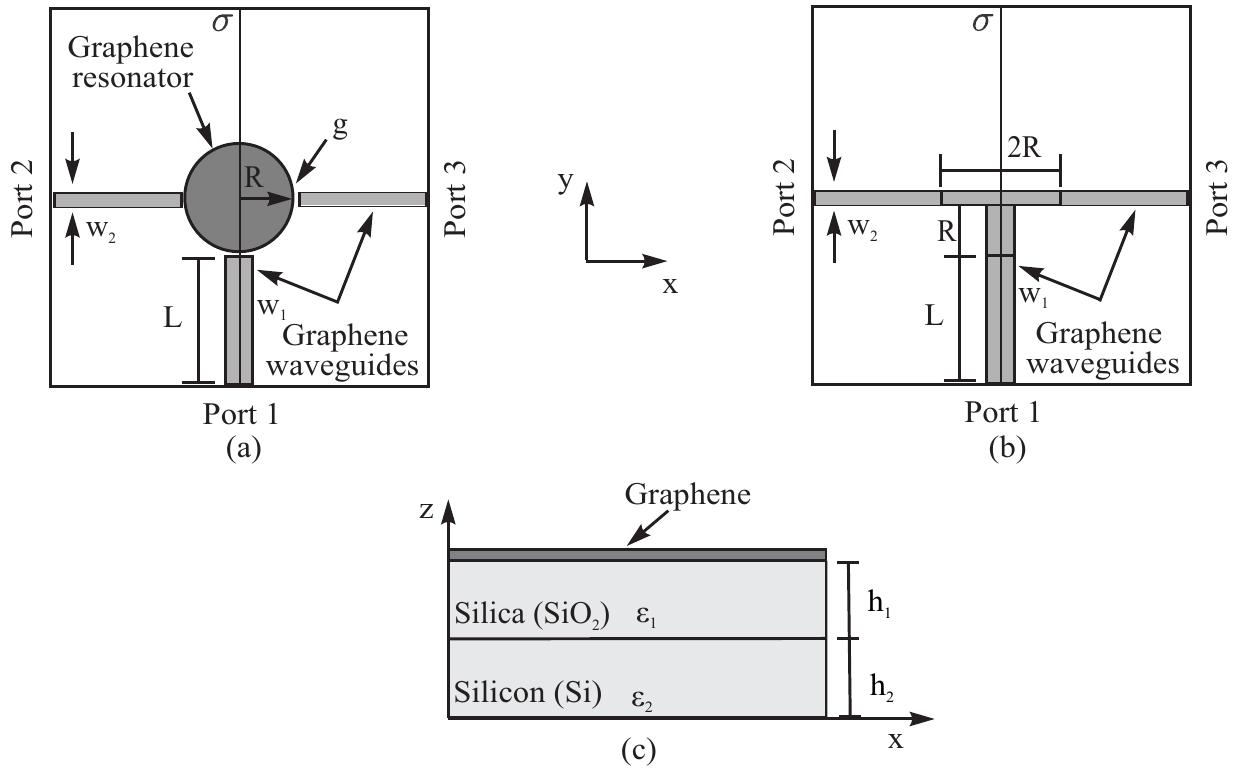}
\caption{Schematic representation of multifunctional graphene components: a) top view of component with resonator,  b) top view of component without resonator, c) side view.}
\label{fig:Fig1}
\end{figure}

The geometries of the components are shown in Fig.\ref{fig:Fig1}, where $\sigma$ is the plane of symmetry of the structure.
In Fig.\ref{fig:Fig1}a, the T-shaped junction with a resonator is presented. It is composed of the graphene circular resonator with radius $R$ = 600 nm and the nanoribbon graphene waveguides with the  propagating fundamental mode. The widths of the nanoribbons are $w_{1}$ = 200 nm, $w_{2}$ = 100 nm and their lengths are $L$ = 1500 nm. The waveguides are frontally coupled to the resonator with a gap $g$ = 5 nm. Graphene elements are deposited on $SiO_{2}$ and $Si$ substrates with thicknesses $h_{1}=h_{2}=2500$ nm and relative permittivity $\varepsilon_{1} = 2.09$ and $\varepsilon_{2} = 11.9$, respectively. 

From the point of view of circuit theory, the input waveguide is connected to the two output ones which are in parallel. Therefore,  we choose the width $w_{2}$ different from $w_{1}$, and this difference was used to improve the impedance matching of the input and the output guides. This allows one to obtain good results for reflection and transmission coefficients \cite{T}. In Fig.\ref{fig:Fig1}b, the central region of the T-structure without resonator has a length in the $x$ direction $2R$ and in the $y$ direction $R$ and the widths of the nanoribbons are $w_{1}$ = 200 nm, $w_{2}$ = 100 nm.

Our task is to calculate and analyse S-parameters of the multifunctional components. Scattering matrix of the reciprocal circuit with the plane of symmetry $\sigma$ shown in Fig.\ref{fig:Fig1} has the following structure:
\begin{equation}
	[S]=\left(\begin{array}{ccc}
	 S_{11} & S_{12} & S_{12}\\ 
	 S_{12} & S_{22} & S_{23}\\
	 S_{12} & S_{23} & S_{22}\\ 
    \end{array} \right),
    \label{eq1}
\end{equation}
where 
\begin{equation}
S_{31} = S_{13} = S_{12}, \; S_{33} = S_{22}, \; S_{32} = S_{23}, \; S_{21} = S_{12}.
\label{eq2}
\end{equation}
\section{Optical Conductivity of Graphene}
For the frequencies from THz to mid-infrared, the Drude model for the local conductivity of graphene mono-layer is frequently used. It is given by the following expression \cite{inter}:
\begin{equation}
\sigma = \frac{2D}{\pi} \frac{i\omega - 1/\tau}{( \omega + i / \tau ) ^2},
\label{eq3}
\end{equation}
where $D = 2\sigma_0 \epsilon_{F} / \hbar$ is the Drude weight, $\sigma_{0}$ is the universal conductivity, $\hbar$ is the reduced Planck constant, $e$ is the electron charge, $\tau=0.9$ ps is the relaxation time and $\omega$ is the frequency of the incident wave. In the following we shall use the Fermi energy of graphene $\epsilon_{F_1}$ = 0.8 eV for the waveguides and $\epsilon_{F_2}$ = 0.6 eV for the disk resonator and $\epsilon_{F_2}$ = 0.8 eV for the central region of the T-shaped device without resonator. Notice, that by electrostatic doping, graphene Fermi level $\epsilon_{F}$ = 1.2 eV can be reached \cite{Khrap}. 

The simulations were fulfilled using the commercial software COMSOL Multiphysics version 5.2.a \cite{comsolSite}, which is based on the method of finite elements. Graphene is a 2D material with the atomic order thickness. However, in our numerical calculus we will consider a graphene monolayer with finite thickness $\Delta$ and conductivity of graphene given by $[\sigma_{v}] = [\sigma_{s}]/\Delta$, where $[\sigma_{s}]$ is is the surface conductivity of graphene \cite {Transformation}. In our simulations, we assume $\Delta = 1$ nm.
\section{Resonator Radius}
An infinite graphene layer placed on a dielectric substrate supports transverse-magnetic (TM) SPP waves with the dispersion relation given by \cite{Livro}:
\begin{equation}
\beta_{spp}=\dfrac{(1+\varepsilon_1)(\omega\hbar)^2}{4\alpha\epsilon_F\hbar c},
\label{eq4}
\end{equation}%
where $\beta_{spp}$ is SPP propagation constant, $\alpha=e^2/(4\pi \varepsilon_0 \hbar c) \approx 0.007$ is the fine-structure constant and $\varepsilon_{1}$ is the dielectric constant of the substrate. 

\begin{figure}[htbp]
\centering
\includegraphics[width=\linewidth]{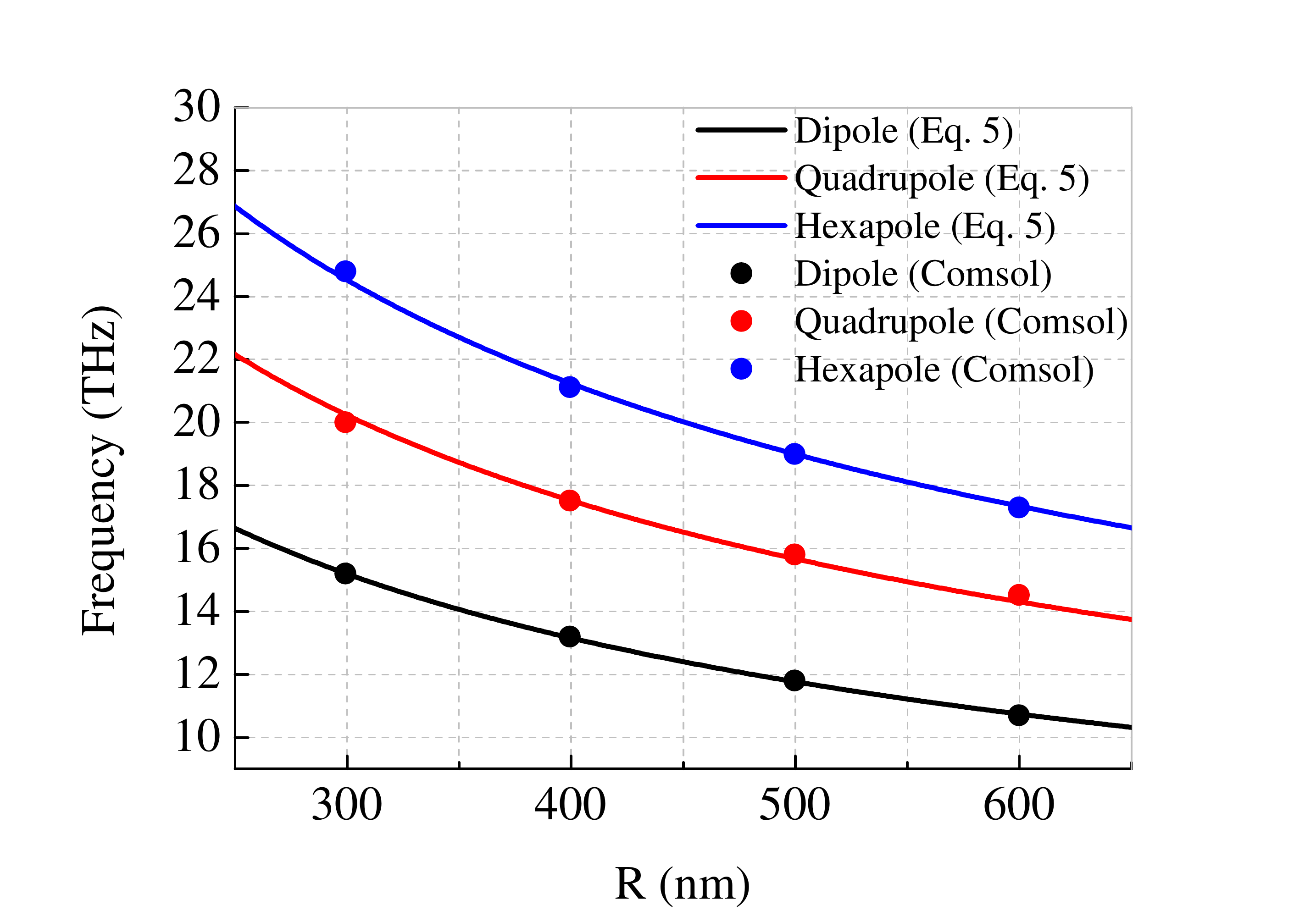}
\caption{Resonant frequency of resonator versus radius $R$, $\epsilon_{F2} = 0.6$ eV.}
\label{fig:RaioEq}
\end{figure}

It is known \cite{borda} that the structure of the fields in the resonator corresponds to edge-guided waves. Thus, one can define the radius $R$ of the resonator with dipole, quadrupole and hexapole mode from the condition of edge-guided mode resonance $2\pi R = n\lambda_{spp}$, where $\lambda_{spp}$ is the wavelength of the SPP mode and n = 1, 2, 3 is the mode number. From the relations $\beta_{spp}=2\pi/\lambda_{spp}$ and $R=n\lambda_{spp}/2\pi$ \cite{Y}, one obtains:
\begin{equation}
R\approx na_nA\dfrac{\epsilon_{F}}{(1+\varepsilon_{1})\omega_{c}^2},
\label{eq5}
\end{equation}%
where $A = 8.3\times10^{40} (kg.m)^{-1}$, $\epsilon_{F}$ is Fermi energy (given in $J$) and $a_{n}$ is the correction coefficient for $n$ mode, $a_{1} = 1.03, a_{2} = 0.97, a_{3} = 0.96$ for the dipole, quadrupole or hexapole mode, respectively. One can see that the radius of the resonator $R$ (given in $m$) depends on the central frequency of the component $\omega_{c}$ (in rad/s), on $\epsilon_{F}$ and $\varepsilon_{1}$. The central frequencies of the components for the modes are increased with decreasing the radius $R$ in accordance with the relation $\omega_{c}\propto1/\sqrt{R}$ as it follows from Eq.\ref{eq5}. Fig.\ref{fig:RaioEq} demonstrates a very good agreement between the analytical and numerical results.
\section{Temporal Coupled Mode Theory} 
Temporal Coupled Mode Theory (TCMT) \cite {Joannopolus, Haus} can be used to calculate the transmission and reflection characteristics of the component. This method is based on some general assumption such as weak coupling of the resonator with external world, linearity of the system and energy conservation. It can be shown that in our case of the component with the plane of symmetry $\sigma$ (see Fig.\ref{fig:Fig1}) and the scattering matrix elements obeying relations Eq.\ref{eq2}, one can consider coupling of the input waveguide with any of the output one separately, i.e. reducing the problem of a three-port to the problem of a two-port. 

The transmission and reflection spectrum of the component can be obtained from the following equations \cite {Joannopolus, Haus}:
\begin{equation}
\frac{da_{j}}{dt} = -i\omega_{j} a_{j} - \frac{a_{j}}{\tau_{r,j}} - \frac{a_{j}}{\tau_{w,j}} + \sqrt{\frac{2}{\tau_{w,j}}} S_{1+},
\label{eq6}
\end{equation}
\begin{equation}
S_{1-} = -S_{1+} + \sqrt{\frac{2}{\tau_{w,j}}} a_{j}, 
\label{eq7}
\end{equation}
where $\frac{1}{\tau_{wj}} = \gamma_{wj}$ are decay rates due to coupling of the resonator with waveguides, $\frac{1}{\tau_{rj}} = \gamma_{rj}$ is the internal losses of the resonator and $\omega_{j}$ is the resonance frequency of the dipole, quadrupole and hexapole modes with j = 1, 2, 3, respectively.  

The transmission coefficients of the component are defined by
\begin{equation} 
S_{31_{j}}=S_{21_{j}}=
\frac{({\gamma_{rj}}-{\gamma_{wj})^2}+(\omega-\omega_{j})^2}{(\omega-\omega_{j})^2 + \biggr({\gamma_{wj}} + {\gamma_{rj}}\biggl)^2}, \; j = 1, 3
\label{eq8}
\end{equation}
\begin{equation} 
S_{21_{2}}=S_{31_{2}}=
\frac{\biggr(2{\gamma_{w2}}\biggl)^2}{(\omega-\omega_{2})^2 + \biggr({\gamma_{w2}} + {\gamma_{r2}}\biggl)^2}.
\label{eq9}
\end{equation}
Summing equations Eq.(\ref{eq8}) and Eq.(\ref{eq9}) gives the parameter $S_{21}$ covering the frequency regions of the dipole, quadrupole and hexapole resonances:
\begin{equation} 
S_{21} = S_{21_{1}} + S_{21_{2}} + S_{21_{3}}.
\label{eq10}
\end{equation}
For the reflection coefficients, one comes to
\begin{equation} 
S_{11_{j}}=
\frac{({\gamma_{rj}}-{\gamma_{wj})^2}+(\omega-\omega_{j})^2}{(\omega-\omega_{j})^2 + \biggr({\gamma_{wj}} + {\gamma_{rj}}\biggl)^2}, \; j = 1, 2, 3
\label{eq11}
\end{equation}
\begin{equation}
S_{11} = S_{11_{1}} + S_{11_{2}} + S_{11_{3}}.
\label{eq12}
\end{equation}
\section{Multifunctional Component with Resonator}
The plane of symmetry $\sigma$ provides in the discussed  T-junctions frequency independent equal division of the input power, i.e. $S_{31} = S_{21}$, see Eq.\ref{eq2}. Below, we consider also filtering properties of the junction with a resonator and, on this basis, we discuss different possibilities of realization of switching function. The component bandwidth is defined by the full width at half maximum (FWHM) of the transmission coefficient curve. 
\subsection{Filtering Properties of T-Component with Resonator}
\begin{figure}[htbp]
\centering
\includegraphics[width=\linewidth]{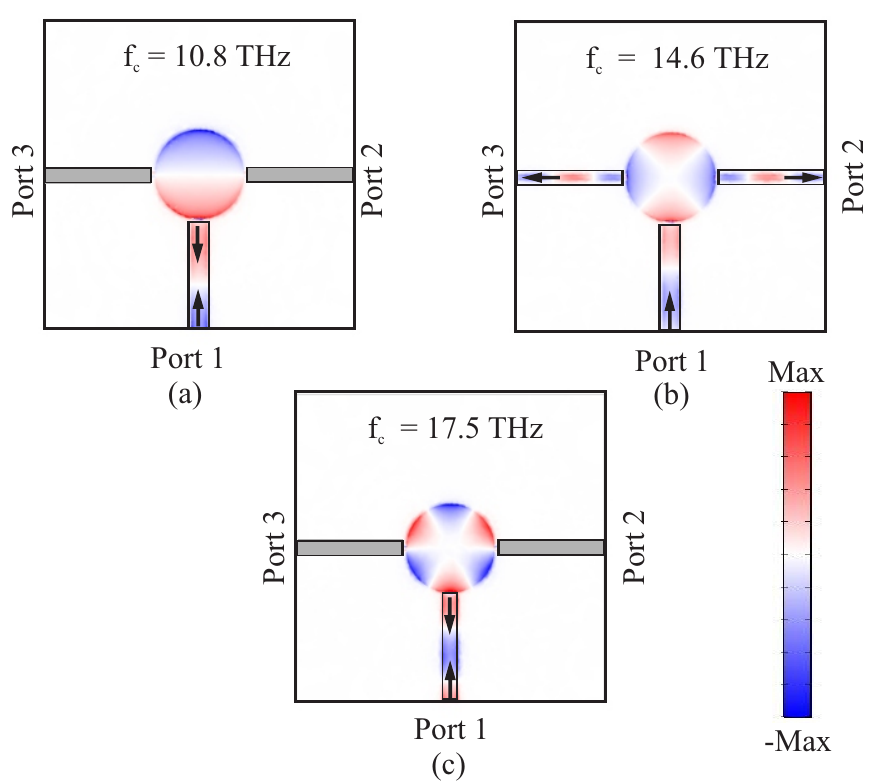}
\caption{$E_{z}$ field distribution, a) for dipole mode, b) for quadrupole mode, c) for hexapole mode, $\epsilon_{F_1}$ = 0.8 eV for the waveguide and $\epsilon_{F_2}$ = 0.6 eV for the resonator.}
\label{fig:Ez}
\end{figure}

\begin{figure}[htbp]
\centering
\includegraphics[width=1\linewidth]{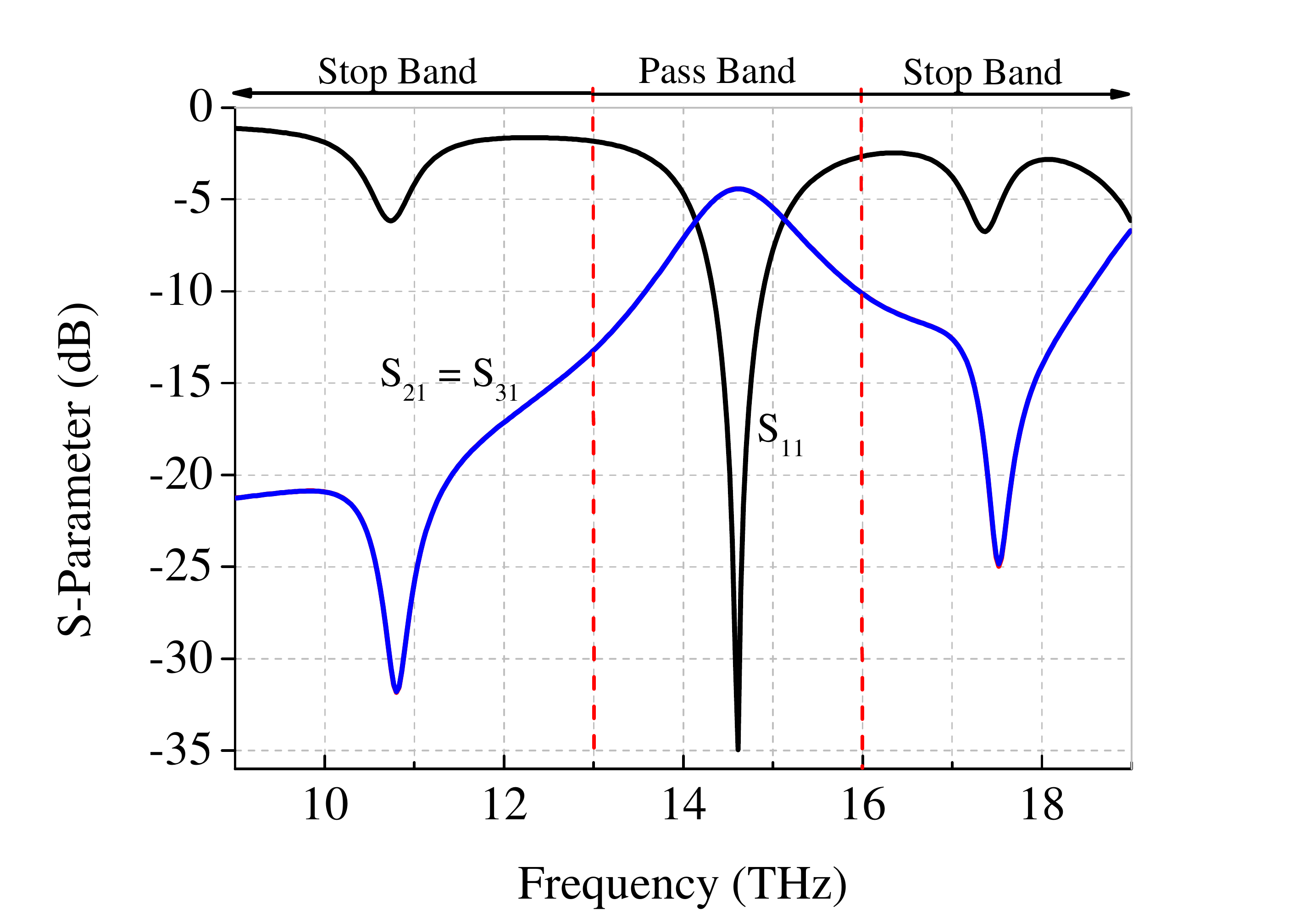}
\caption{Frequency responses of multifunctional device, $\epsilon_{F_1}$ = 0.8 eV and $\epsilon_{F_2}$ = 0.6 eV.}
\label{fig:FigDisp}
\end{figure}

The input electromagnetic wave in port 1 produces a dipole, or quadrupole or hexapole resonance in the circular graphene resonator (Fig.\ref{fig:Ez}a,b,c). 
The quadrupole resonance gives band-pass filter. For the dipole and hexapole resonances, ports 2 and 3 are isolated from the input one. This is explained by coincidence of the nodal planes of the dipole and hexapole modes with the plane of symmetry of the output ports. Thus, these frequency regions correspond to the band-stop filters. The frequency response for the resonant modes are shown in Fig.\ref{fig:FigDisp}. It can be seen that at the dipole resonance frequency of 10.8 THz, the isolation level of ports 2 and 3 are around -31.8 dB. For hexapole mode we have the resonant frequency 17.5 THz and the isolation level of ports 2 and 3 is -25 dB. The reflection coefficients for the respective  frequencies are -6.1 dB and -6.7 dB.
\subsection{Control of Resonances by Fermi Energy}
\begin{figure}[htbp]
\centering
\includegraphics[width=1\linewidth]{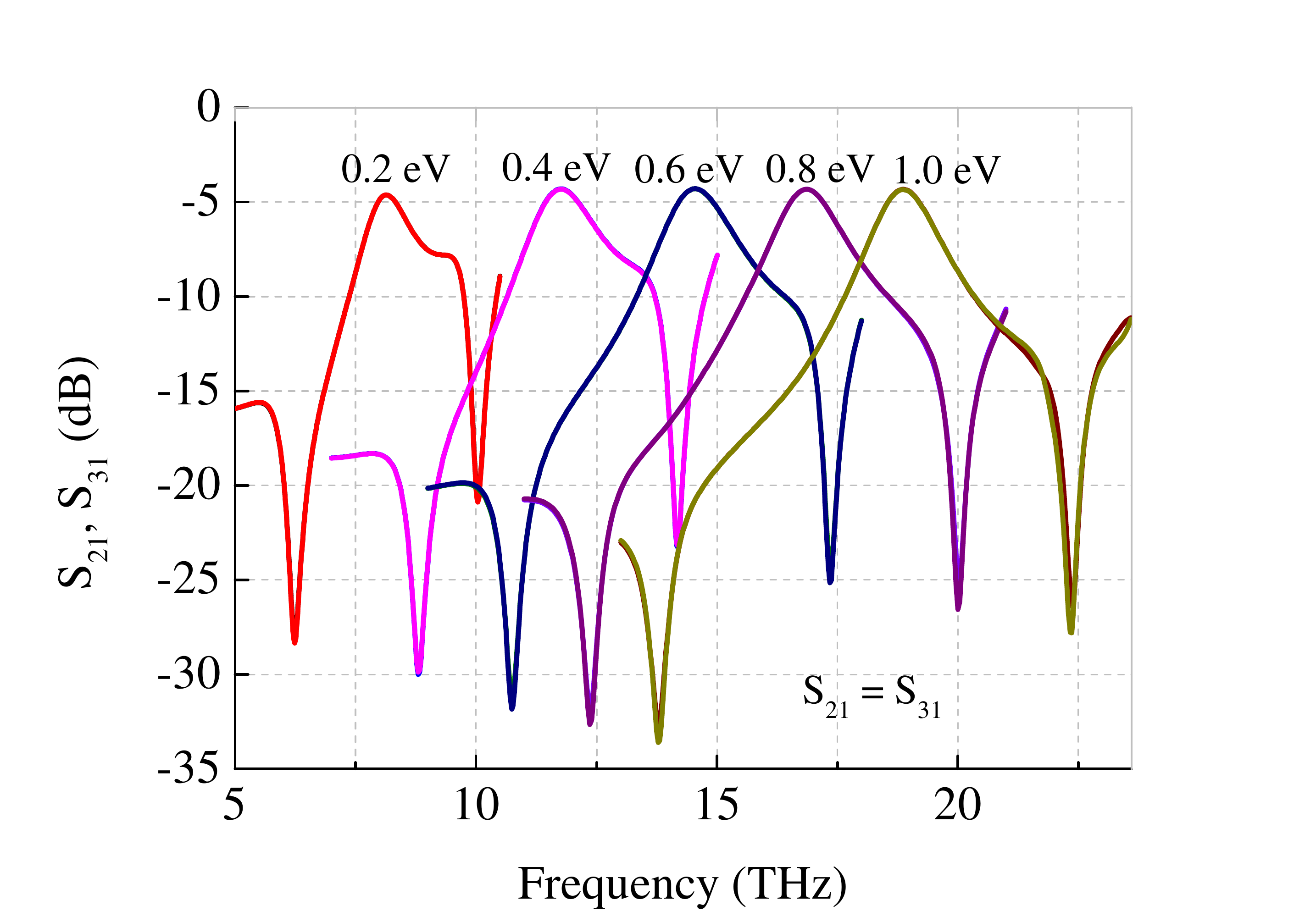}
\caption{Frequency responses of multifunctional component for different values of $\epsilon_{F_1}=0.8$ eV and $\epsilon_{F_2}$.}
\label{fig:Mi}
\end{figure}

The influence of the Fermi energy on the resonance frequencies of the structure was investigated. Fermi energy the waveguide was fixed at $\epsilon_{F_1}$ = 0.8 eV and  $\epsilon_{F_2}$ of the resonator was varied from 0.2 eV to 1 eV. The corresponding frequency responses are plotted in Fig.\ref{fig:Mi}.
The increase of the Fermi energy in the resonator shifts the resonant frequency to higher values. With change of $\epsilon_{F_2}$ between 0.2 eV and 1.0 eV, the resonant frequency of the dipole mode ranges from 6.2 THz to 13.8 THz for quadrupole mode from 8 THz to 18.8 THz and for hexapole mode from 10 THz to 22.3 THz.
\subsection{Possible Switching regimes}
The multifunctional component is shown in Fig.\ref{fig:Fig1}a. The physical principle of its functioning is based on the propagation of SPP waves in the graphene waveguides, which excite resonances of in the resonator. This component can operate as a controllable  filter, a power divider or switch, depending on the Fermi energy of graphene resonator. 
For resonance of the quadrupole mode, the transmitted power is divided equally between ports 2 and 3, therefore one can use the device as a power divider and band-pass filter, and this is seen in  the field distribution of the electric field component $E_{z}$ shown in Fig.\ref{fig:Ez}b. For the quadrupole mode, the insertion losses in ports 2 and 3 is -4.3 dB at the central frequency 14.6 THz and reflection coefficient is -35 dB. The frequency response for quadrupole mode is presented in Fig.\ref{fig:FigDisp}.   

To use the resonator component as a switch, we have three possibilities, namely, to move the dipole mode resonance to the quadrupole mode region ((D+Q) switching), to move the hexapole mode to the quadrupole mode region ((Q+H) switching), or to put the
the Fermi energy of graphene resonator equals to zero. 

The frequency changes of resonant modes can be fulfilled by applying an external voltage to the graphene disc layer, changing its Fermi energy and shifting the quadrupole resonant mode to higher ((D+Q) switching) or the hexapole resonant mode to lower frequencies ((Q+H) switching).
\subsubsection{(D+Q) Switching}
\begin{figure}[htbp]
\centering
\includegraphics[width=1\linewidth]{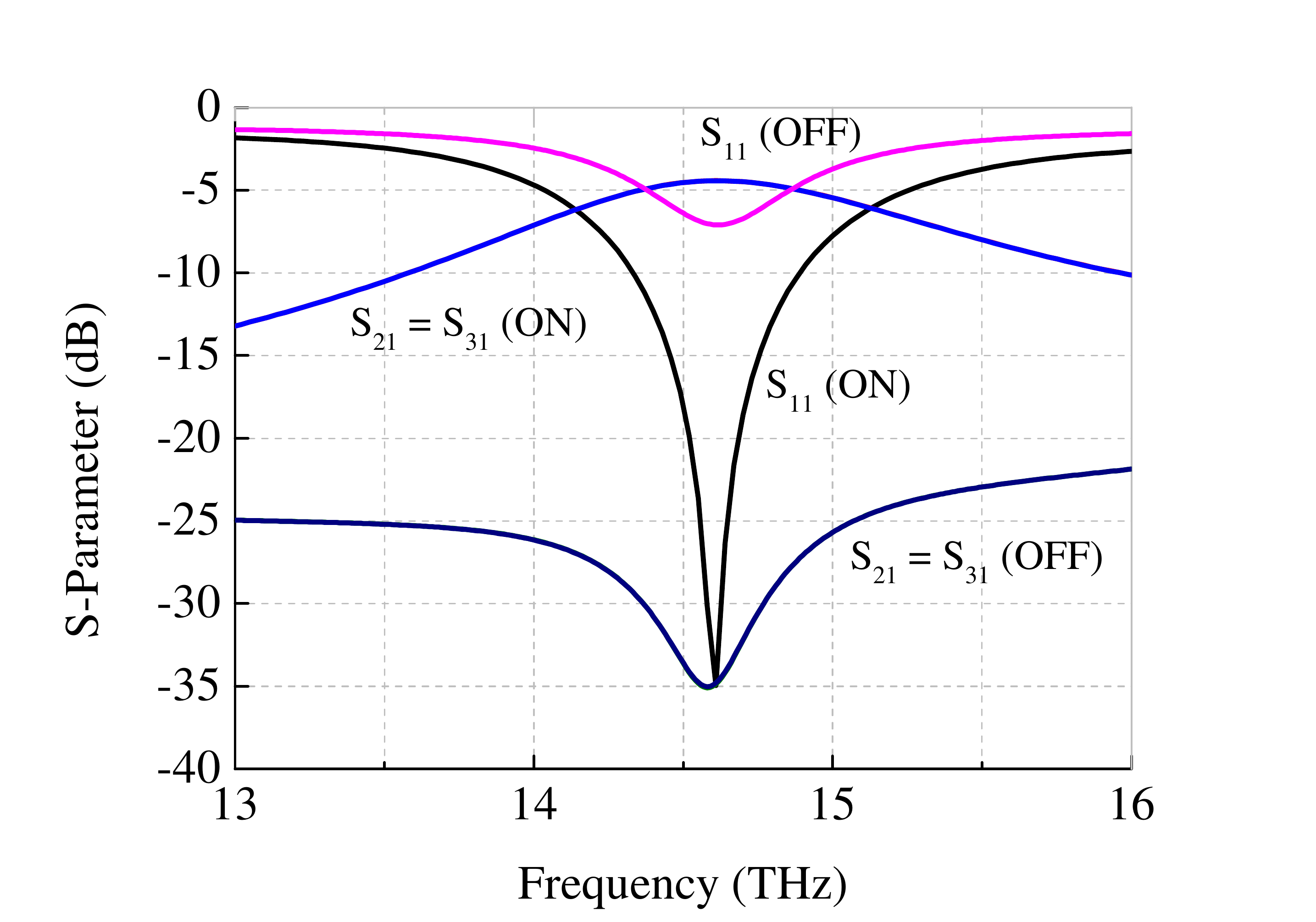}
\caption{Frequency responses of (D+Q) switch in ON ($\epsilon_{F_1}$ = 0.8 eV and $\epsilon_{F_2}$ = 0.6 eV ) and OFF ($\epsilon_{F_1}$ = 0.8 eV and $\epsilon_{F_2}$ = 1.11 eV) states.}
\label{fig:ONOFF111}
\end{figure}

Note that for the ON state of the configuration (D+Q), we have the losses of -4.3 dB for ports 2 and 3, and the switch operating region is 9.5$\%$. The reflection coefficient at the central frequency is about -35 dB. For the OFF state, we have the losses about -35 dB in port 2 and in port 3 at the central frequency  (see Fig.\ref{fig:ONOFF111}). 
\subsubsection{(Q+H) Switching}
\begin{figure}[htbp]
\centering
\includegraphics[width=1\linewidth]{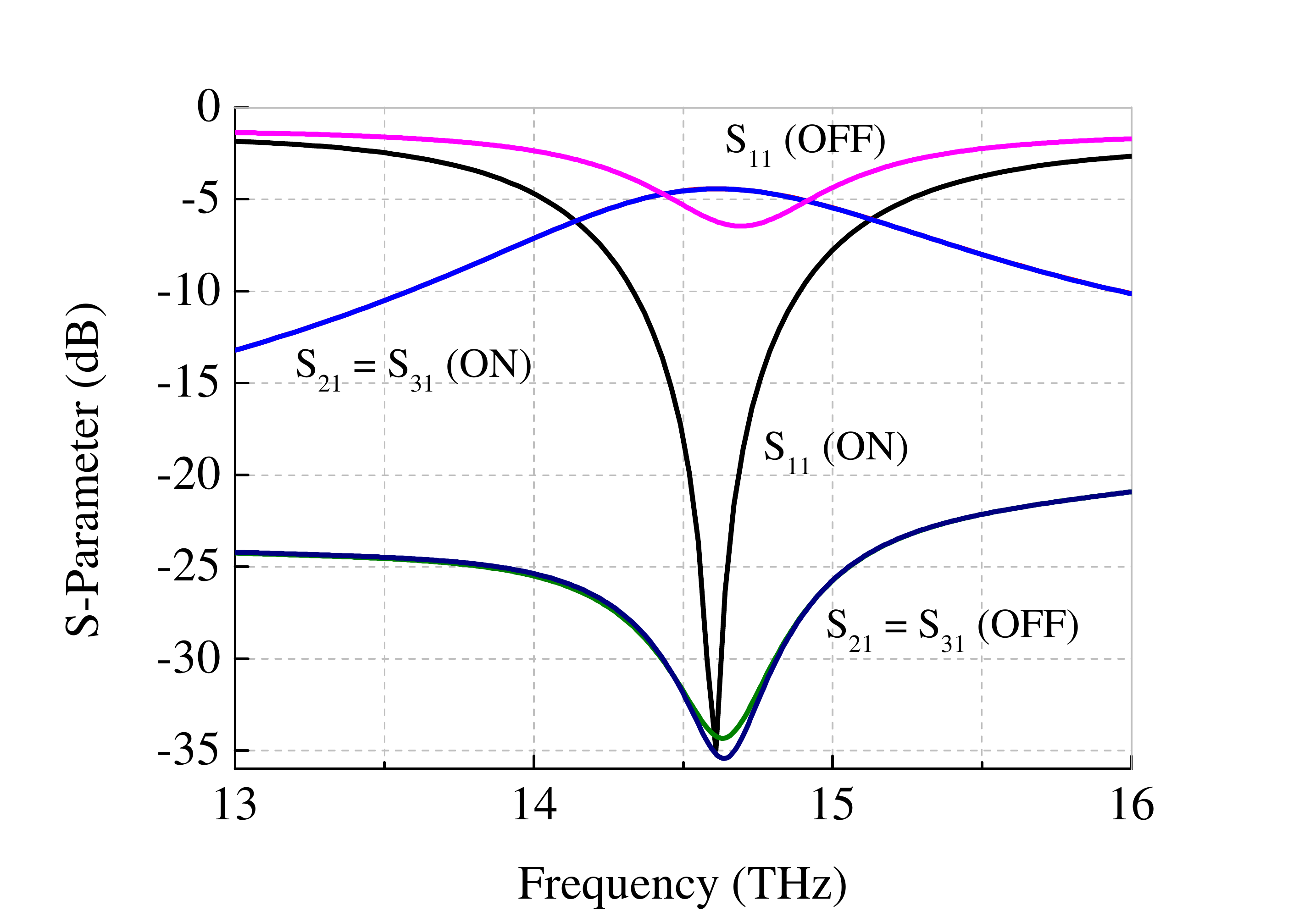}
\caption{Frequency responses of (Q+H) switch in ON ($\epsilon_{F_1}$ = 0.8 eV and $\epsilon_{F_2}$ = 0.6 eV) and OFF ($\epsilon_{F_1}$ = 0.8 eV and $\epsilon_{F_2}$ = 0.42 eV) states.}
\label{fig:ONOFF42}
\end{figure}

For configuration (D+H) in ON state, we have the losses -4.3 dB for ports 2 and 3, and the reflection coefficient of -35 dB. For the OFF state, we have the losses of -35.5 dB for ports 2 and -34.3 dB for port 3 and the reflection coefficient of -6.5 dB, with 9.5$\%$ bandwidth (see Fig.\ref{fig:ONOFF42}). 
%
\subsubsection{Switching by Zero Fermi Energy}
\begin{figure}[htbp]
\centering
\includegraphics[width=1\linewidth]{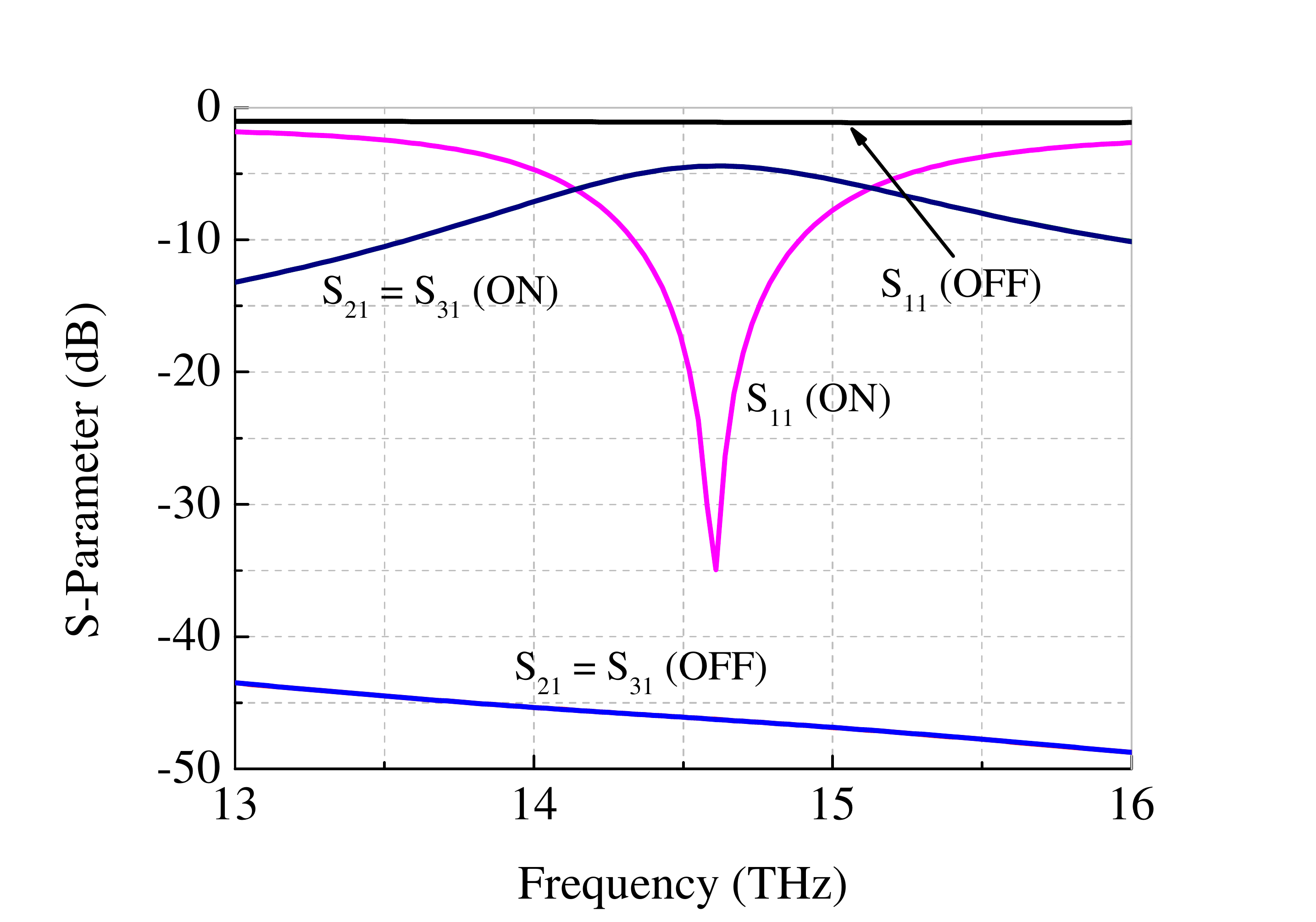}
\caption{Frequency responses of (Q+D) switch in ON ($\epsilon_{F_1}$ = 0.8 eV and $\epsilon_{F_2}$ = 0.6 eV) and OFF ($\epsilon_{F_1}$ = 0.8 eV and $\epsilon_{F_2}$ = 0 eV) states.}
\label{fig:ONOFF0}
\end{figure}

For the OFF state, one can also choose $\epsilon_{F_2}$ = 0. In this case, the conductivity of the graphene is very small, and the SPP modes are not excited in the resonator. The most part of the incident power is reflected from the resonator. The reflection is on the level of -1 dB and the isolation of ports 2 and 3 is -43 dB. For the ON state, the results are the same as those of the (D+Q). The frequency responses for the switch are shown in  Fig.\ref{fig:ONOFF0}.
\subsection{Comsol versus TCMT} 
\begin{figure}[htbp]
\centering
\includegraphics[width=\linewidth]{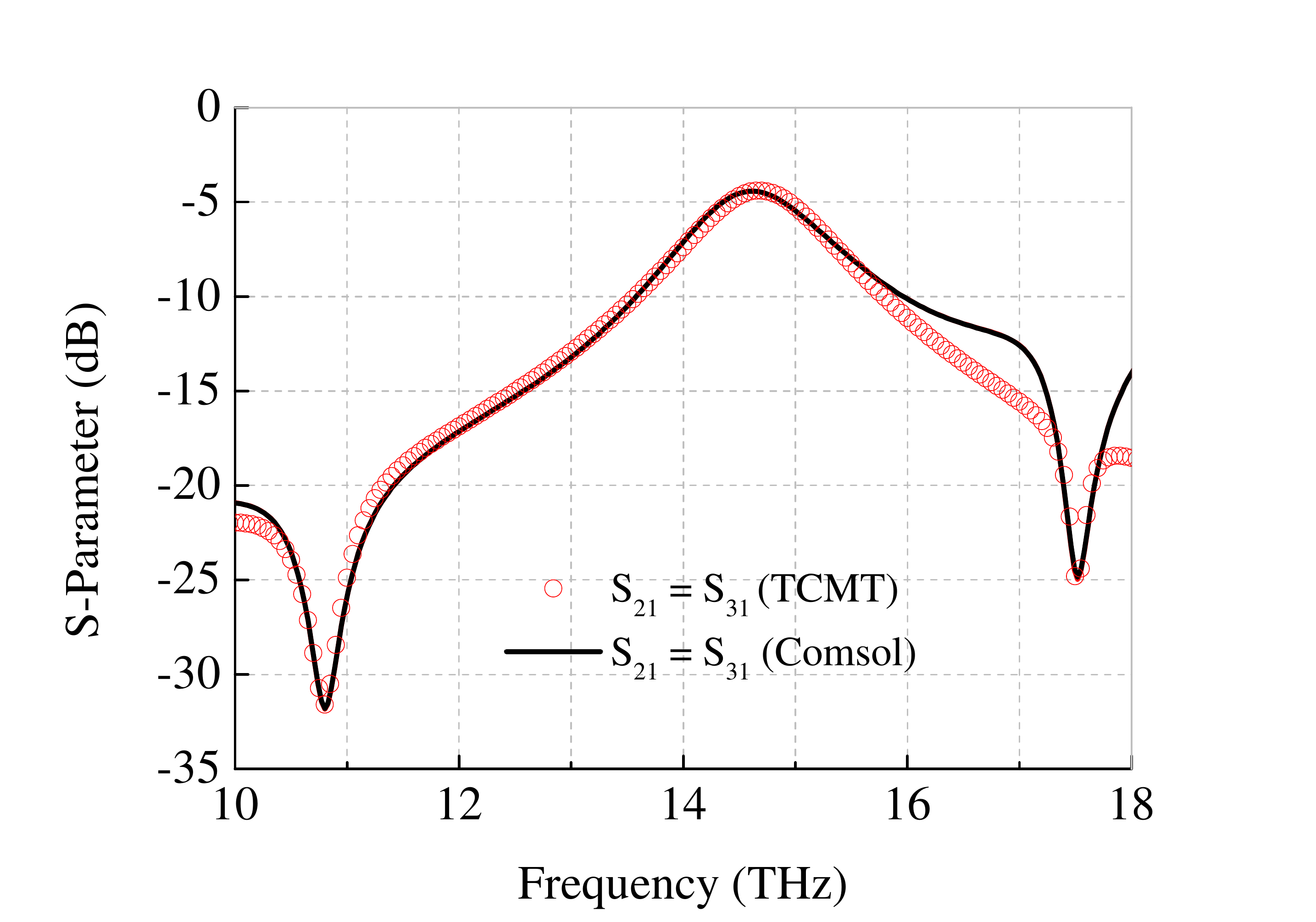}
\caption{Frequency responses of multifunctional device simulated by Comsol (continuous lines) and by TCMT (circles), $\epsilon_{F_1}$ = 0.8 eV and $\epsilon_{F_2}$ = 0.6 eV.}
\label{fig:TCMT1}
\end{figure}

For TCMT calculus of the resonance frequencies and the decay rates were obtained from Comsol simulations. The calculated parameters are as follows: $\omega_{1} = 67.86$ $10^{12}$ rad/s, $\omega_{2} = 91.73$ $10^{12}$ rad/s, $\omega_{3} = 109.95$ $10^{12}$ rad/s, $\gamma_{w1}\cong 1.5 $ THz, $\gamma_{i1} \cong 0.9$ THz, $\gamma_{w2}\cong 1.303$ THz, $\gamma_{i2}\cong 2.997$ THz, $\gamma_{w3}\cong 0.58621$ THz, $\gamma_{i3}\cong 0.26379$ THz to the dipole, quadrupole and hexapole modes, respectively. Fig.\ref{fig:TCMT1} shows the numerical results obtained by Comsol and the analytical results obtained by TCMT. One can notice a good agreement between the two methods.
\section{Two-Functional Component without Resonator} 
Finally, we shall discuss a T-shaped component without resonator  shown in Fig.\ref{fig:Fig1}b. In case of the optimized widths  $w_{1}$ = 200 nm and $w_{2}$ = 100 nm, SPP waves propagating in the input waveguide and reaching the central region of the device are equally divided to the output guides (ON state). So, the device can function as a power divider. In order for this component to function as a switch, one can change the Fermi energy of its central region to zero. In this case, the wave is reflected and does not reach the output ports (OFF state). 

\begin{figure}[htbp]
\centering
\includegraphics[width=\linewidth]{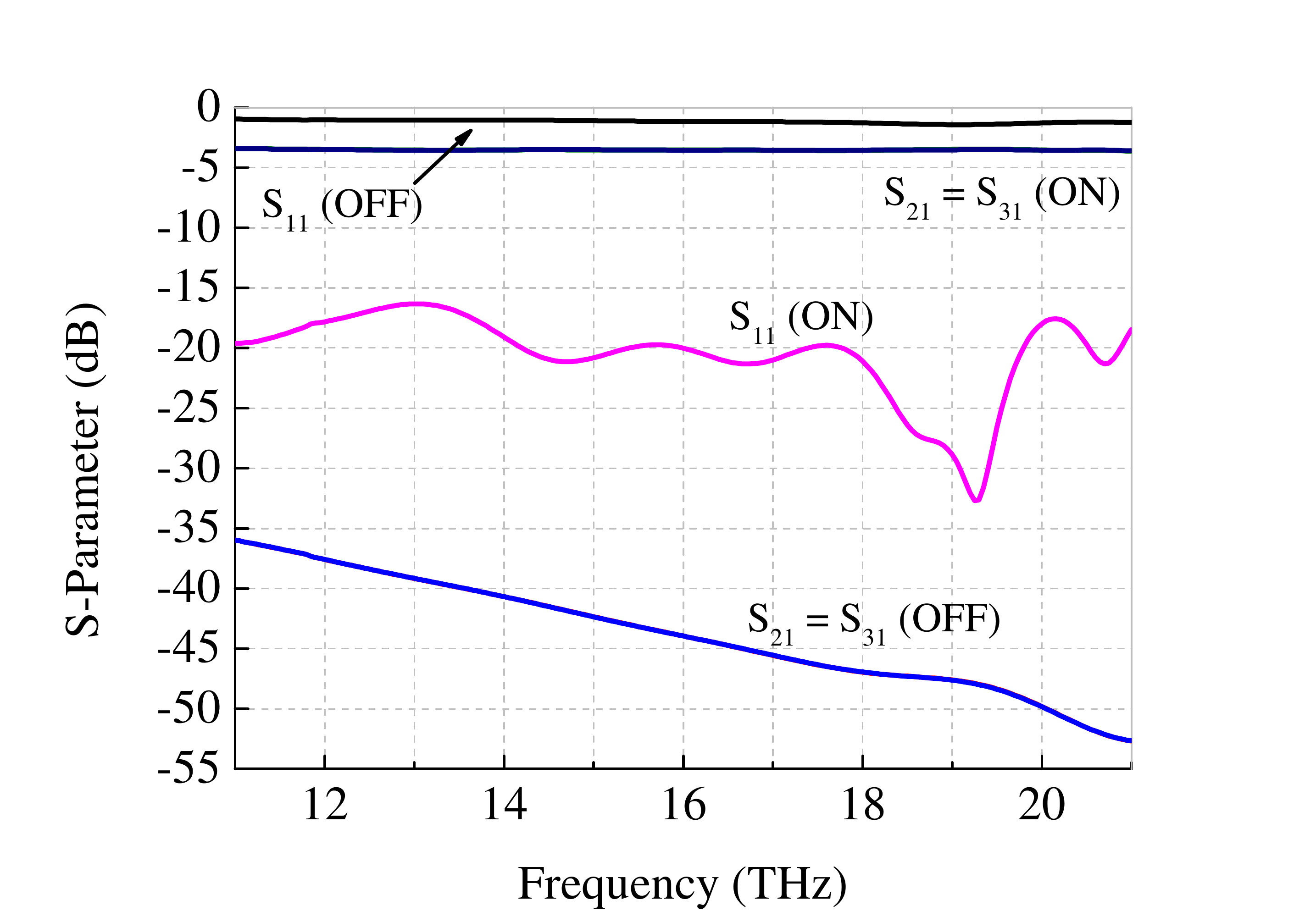}
\caption{Frequency response for T-shaped component without resonator.}
\label{fig:ONOFF08}
\end{figure}

\begin{figure}[htbp]
\centering
\includegraphics[width=\linewidth]{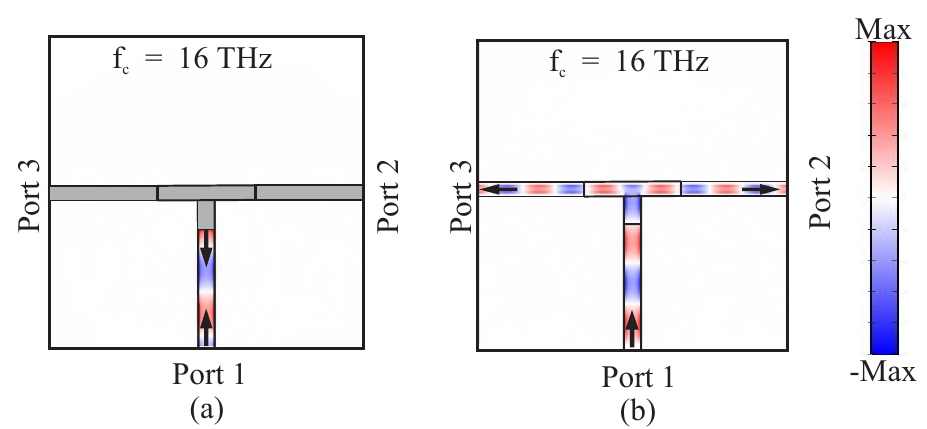}
\caption{$E_z$ field distribution in T-shape component without resonator, a) $\epsilon_{F_1}$ = 0.8 eV and $\epsilon_{F_2}$ = 0.0 eV (OFF state), b) $\epsilon_{F_1}$ = 0.8 eV and $\epsilon_{F_2}$ = 0.8 eV (ON state).}
\label{fig:T}
\end{figure}

We choose the Fermi energy of the waveguides $\epsilon_{F_1}$ = 0.8 eV. If we put the Fermi energy in the central region equal to 
$\epsilon_{F_2}$ = 0 eV (OFF state), the reflection coefficient is -1 dB and the losses at -36 dB level for ports 2 and 3. Changing the Fermi energy of the central region of T-junction to 
$\epsilon_{F_2}$ = 0.8 eV, one comes to the state ON. In this case, the losses for ports 2 and 3 are the level of -3.5 dB and the reflection coefficient around -20 dB. Fig.\ref{fig:ONOFF08} and Fig.\ref{fig:T} show the frequency responses of the device and the $E_{z}$ field distribution in the waveguides. 
Thus, by this example we have demonstrated a possibility to construct a wideband divider-switch in THz and far-infrared regions.

In the discussed above two components, their control (switching and dislocation of the frequency band) is provided by Fermi energy by application of a DC voltage (i.e. by external electric field normal to the graphene plane). As a result, this leads to a change of the graphene conductivity. We have shown in Fig.\ref{fig:Fig1} the graphen-substrate structures without details of the control system. In practice, the DC voltage can be applied between the graphene sheet and a very thin polysilicon layer with relatively high conductivity. This layer placed close to graphene is used as a gate electrode \cite{Gomez}, \cite{Yan}, \cite{Fus}.
\section{Conclusion}
We have analyzed two  graphene-based components for THz and far-infrared regions. Both of them have T-shaped  geometry. The first one can substitute three common elements. It contains a circular resonator. The other component is without resonator. The resonator component can work with one of the three resonant modes: dipole, quadrupole and hexapole ones. It can function as a dynamically controllable filter, power divider and a switch, thus, substituting three common elements. Besides it permits a very wideband control of its central frequency.  

We have analyzed dependence of the resonant frequencies on the resonator radius. Our analysis demonstrates also that the resonant frequencies of the component can be adjusted by changing graphene Fermi energy, which enables the dynamical control of the central frequency in a broad frequency region. The component without resonator can function as a broadband divider and a switch. The presented components are very simple and compact. They have  a relatively low insertion loss and good matching in the regime ON and high isolation of the output ports in the regime OFF. The suggested components can be used in future THz and far-infrared devices and systems.
\section*{Funding}
This work was supported by the Brazilian agency National Counsel of Technological and Scientific Development CNPq and Pro-Rectory of Research and Post-Graduation PROPESP/UFPA.
\section*{Disclosures}
The authors declare that there are no conflicts
of interest related to this paper.
%


%

\begin{thebibliography}{1}
\newcommand{\enquote}[1]{``#1''}




\bibitem {Giani}
V.~Dimitriev, G.~Portela, L.~Martins and D.~Zimmer, \enquote{Temporal coupled-mode theory of eletromagnetic components with magnetic symmetry,} {\it{IEEE Trans. Microw. Theory Tech.}} \textbf{66}, 1165--1171 (2018).


\bibitem{L}
A.~Dolatabady and N.~Granpayeh, \enquote{L-shaped filter, mode separator and power divider based on plasmonic waveguides with nanocavity resonators,} {\it{IET Optoelectronics}} \textbf{9}, 289--293 (2015).

\bibitem{Design}
Y. Ye et al., \enquote{Design of a Novel Plasmonic Splitter With Variable Transmissions and Selectable Channels,} {\it{IEEE Trans. Nanotec.}} \textbf{18}, 617--625 (2019).

\bibitem{Numerical}
D.~C.~Tee, N.~Tamchek, Y.~G.~Shee, and F.~R.~Mahamd~Adikan, \enquote{Numerical investigation on cascaded 1$\times$3 photonic crystal power splitter based on asymmetric and symmetric 1$\times$2 photonic crystal splitters designed with flexible structural defects,} {\it{Opt. Express}} \textbf{22}, 24241--24255 (2014).

\bibitem{Beam}
Zheng,~G., Xu,~L., Chen,~Y. et al., \enquote{Beam filter and splitter based on surface plasmon propagation in ring metal heterowaveguide,} {\it{Pramana - J. Phys.}} \textbf{83}, 995--1002 (2014).

\bibitem{splitter}
Y.~Guo, L.~Yan, W.~Pan, B.~Luo, K.~Wen, Z.~Guo, H.~Li, and X.~Luo, \enquote {A plasmonic splitter based on slot cavity,} {\it{Opt. Express}} \textbf{19}, 13831--13838 (2011).

\bibitem{Geim}
A.~K.~Geim, and K.~S.~Novoselov, \enquote{The Rise Of Graphene,} {\it{Nat. Mater.}}, \textbf{6}, 183--191 (2007).

\bibitem{Slot}
S.~Asgari and N.~Granpayeh, \enquote{Applications of Tunable Nanoscale Midinfrared Graphene Based Slot Cavity in Nanophotonic Integrated Circuits,} {\it{IEEE Trans. Nanotec.}} \textbf{17},533--542 (2018).

\bibitem{Bandpass}
H.~Zhuang, F.~Kong, K.~Li, and S.~Sheng, \enquote{Plasmonic bandpass filter based on graphene nanoribbon,} {\it{Appl. Opt.}} \textbf{54}, 2558--2564 (2015).

\bibitem{filtro}
Alireza Tavousi, Mohammad Ali Mansouri-Birjandi, and Morteza Janfaza, \enquote{Optoelectronic application of graphene nanoribbon for mid-infrared bandpass filtering,} {\it{Appl. Opt.}} \textbf{57}, 5800--5805 (2018).
 
\bibitem{chave}
Morteza Janfaza, Mohammad Ali Mansouri-Birjandi and Alireza Tavousi, \enquote{Dynamic switching between single and double plasmon induced reflection through graphene nanoribbons based structure,} {\it{Mater. Res. Express}} \textbf{5}, 115022 (2018).

\bibitem{sensor}
Morteza Janfaza, Mohammad Ali Mansouri-Birjandi, Alireza Tavousi, \enquote{Tunable plasmon-induced reflection based on graphene nanoribbon Fabry-Perot resonator and nanodisks,} {\it{Optical Materials}} \textbf{84}, 675--680 (2018).

\bibitem{modulator}
Baohu Huang, Weibing Lu, Zhenguo Liu, and Siping Gao, \enquote{Low-energy high-speed plasmonic enhanced modulator using graphene,} {\it{Opt. Express}} \textbf{26}, 7358--7367 (2018).

\bibitem{divisor}
Bing Wang, Xiang Zhang, Xiaocong Yuan, and Jinghua Teng, \enquote{Optical coupling of surface plasmons between graphene sheets,} {\it{Appl. Phys. Lett.}} \textbf{100}, 131111 (2012).

\bibitem{Graphene}
Z.~Su, X.~Chen, J.~Yin, and X.~Zhao, \enquote{Graphene-based terahertz metasurface with tunable spectrum splitting,} {\it{Opt. Lett.}} \textbf{41} 3799--3802 (2016).

\bibitem{ultrashort}
Huang, C., Sun, T., \enquote{Numerical simulations of tunable ultrashort power splitters based on slotted multimode interference couplers,} {\it{Sci. Rep.}} \textbf{9}, 12756 (2019).
 
\bibitem{wagner}
V.~Dmitriev and W.~Castro, \enquote{Dynamically controllable graphene terahertz splitters with nonreciprocal properties,} {\it{Appl. Opt.}} \textbf{58}, 6513--6518 (2019).

\bibitem{ultra}
J.~Yang, H.~Xin, Y.~Han, D.~Chen, J.~Zhang, J.~Huang, and Z.~Zhang, \enquote{Ultra-compact beam splitter and filter based on a graphene plasmon waveguide,} {\it{Appl. Opt.}} \textbf{56}, 9814--9821 (2017).

\bibitem{Four}
W.~Su, \enquote{A Four-Port Ultra-Compact Terahertz Splitting Filter Based on Graphene Nanoribbon,} {\it{IEEE Photon. Technol. Lett.}} \textbf{31}, 86--89 (2019).

\bibitem{Nikitin}
A. Y. Nikitin, P. Alonso González, S. Vélez, S. Maste, A. Centeno, A. Pesquera, A. Zurutuza, F. Casanova, L. E. Hueso, F. H. L. Koppens, R. Hillenbrand, \enquote{Real-space mapping of tailored sheet and edge plasmons in graphene nanoresonators,} {\it{Nat. Photon.}} \textbf{10}, 239--243 (2016).

\bibitem{Zhang}
L.~Zhang, J.~Yang, X.~Fu, and M.~Zhang, \enquote{Graphene disk as an ultra compact ring resonator based on edge propagating plasmons,} {\it{Appl. Phys. Lett.}} \textbf{103}, 163114 (2013).

\bibitem{filter}
S.~Sheng, K.~Li, F.~Kong, H.~Zhuang, \enquote{Analysis of a tunable band-pass plasmonic filter based on graphene nanodisk resonator,} {\it{Opt. Commun.}} \textbf{336}, 189--196 (2015).

\bibitem{90}
H.~J.~Li, L.~L.~Wang, Z.~R.~Huang, B.~Sun, X.~Zhai, X.~F.~Li, \enquote{Simulations of multi-functional optical devices based on a sharp $90^\circ$ bending graphene parallel pair,} {\it{J. Opt.}} \textbf{16}, 015004 (2014).

\bibitem{Mid}
K.~J.~A.~Ooi, H.~S.~Chu, L.~K.~Ang, and P.~Bai, \enquote{Mid-infrared active graphene nanoribbon plasmonic waveguide devices,} {\it{J. Opt. Soc. Am. B}} \textbf{30}, 3111--3116 (2013).

\bibitem{T}
X.~Zhu, W.~Yan, N. A.~Mortensen, and S.~Xiao, \enquote{Bends and splitters in graphene nanoribbon waveguides,} {\it{Opt. Express}} \textbf{21}, 3486--3491 (2013).

\bibitem{inter}
Y.~V.~Bludov, A.~Ferreira, N.~M.~R.~Peres, and M.~I.~Vasilevskiy,
\enquote{A primer on surface plasmon-polaritons in graphene,} {\it{Int. J. Mod. Phys. B}} \textbf{27}, 1341001 (2013).

\bibitem{Khrap}  
Khrapach, I. et al., \enquote{Novel Highly Conductive and Transparent Graphene-Based Conductors,} {\it{Adv. Mater}} \textbf{24}, 2844--2849 (2012).

\bibitem{comsolSite}
\url{http://www.comsol.com.br}. 

\bibitem{Transformation}
 A.~Vakil, N.~Engheta, \enquote{Transformation optics using graphene,} {\it{Science}} \textbf{332}, 1291--1294 (2011).

\bibitem{Livro} 
P.~A.~D.~Gonçalves and N.~M.~R.~Peres, \enquote{An Introduction to Graphene Plasmonics,} (World Scientific, 2016).

\bibitem{borda}
J.~Wang, W.~B.~Lu, X.~B.~Li, Z.~H.~Ni and T.~Qiu, \enquote{Graphene plasmon guided along a nanoribbon coupled with a nanoring,} {\it{J. Phys. D, Appl. Phys.}} \textbf{47}, 135106 (2014).

\bibitem{Y}
V. Dmitriev and W. Castro, \enquote{Dynamically controllable terahertz graphene Y-circulator,}{\it{IEEE Trans. Magnetics}} \textbf{55}, 4001712 (2019).

\bibitem{Joannopolus} 
J.~D.~Joannopolus, S.~G.~Johnson, J.~N.~Winn, and R.~D.~Meade \enquote{Photonic Crystals,} {\it{Princeton University Press}} (New Jersey, 2007).

\bibitem{Haus} 
H.~A.~Haus, and W.~Huang, \enquote {Coupled-Mode Theory,} {\it{Proceedings of the IEEE}} \textbf{79}, 1505--1518 (1991).
\bibitem{Gomez}
J.S. Gomez-Diaz, C. Moldovan, S. Capdevila, J. Romeu, L.S. Bernard, A. Magrez, A.M. Ionescu, J.Perruisseau-Carrier, \enquote{Self-biased reconfigurable graphene stacks for terahertz plasmonics,} {\it{Nat. Commun.}} \textbf{6}, 6334 (2015). 
\bibitem{Yan}
Y. Li et al., \enquote{Graphene-Based Floating-Gate Nonvolatile Optical Switch,} {\it{IEEE Photon. Technol. Lett.}} \textbf{28}, 284--287 (2016).
\bibitem{Fus} 
Fuscaldo, W., Burghignoli, P., Baccarelli, P., Galli, A., \enquote{Graphene Fabry-Perot cavity leaky-wave antennas: Plasmonic versus nonplasmonic solutions,} {\it{IEEE Trans. Antennas Propag.}} \textbf{65}, 1651--1660 (2017).
\end{thebibliography}
\end{document}